# Twist-Dependent Anisotropic Thermal Conductivity in Homogeneous MoS$_2$ Stacks

Wenwu Jiang[1#], Ting Liang[2#], Jianbin Xu[2], Wengen Ouyang[1*]

[1]Department of Engineering Mechanics, School of Civil Engineering, Wuhan University, Wuhan, Hubei 430072, China

[2]Department of Electronic Engineering and Materials Science and Technology Research Center, The Chinese University of Hong Kong, Shatin, N.T., Hong Kong SAR, 999077, P. R. China

\#: These authors contributed equally.

*E-mail: w.g.ouyang@whu.edu.cn (W. Ouyang)

**Abstract**

Thermal transport property of homogeneous twisted molybdenum disulfide ($MoS_2$) is investigated using non-equilibrium molecular dynamics simulations with the state-of-art force fields. The simulation results demonstrate that the cross-plane thermal conductivity strongly depends on the interfacial twist angle, while it has only a minor effect on the in-plane thermal conductivity, exhibiting a highly anisotropic nature. A frequency-decomposed phonon analysis showed that both the cross-plane and in-plane thermal conductivity of $MoS_2$ are dominated by the low-frequency phonons below 15 THz. As the interfacial twist angle increases, these low-frequency phonons significantly attenuate the phonon transport across the interface, leading to impeded cross-plane thermal transport. However, the in-plane phonon transport is almost unaffected, which allows for maintaining high in-plane thermal conductivity. Additionally, our study revealed the strong size dependence for both cross-plane and in-plane thermal conductivities due to the low-frequency phonons of $MoS_2$. The maximum in-plane to cross-plane thermal anisotropy ratio is estimated as 400 for twisted $MoS_2$ from our simulation, which is in the same order of magnitude as recent experimental results (~900). Our study highlights the potential of twist engineering as a tool for tailoring the thermal transport properties of layered materials.

**Introduction**

Twist engineering, in which one layer of a bilayer system is twisted relative to another, provides a plentiful degree of freedom for manipulating the properties of van der Waals two-dimensional (2D) materials, such as electrical,[1-3] mechanical,[4, 5] and tribological properties.[6-8] The flourishing of twisted systems even triggered the emergence of "Twistronics".[9-11] Motivated by "Twistronics", the effect of the twist angle on the thermal transport of 2D materials has also been extensively studied, as thermal issues are vital to the lifetime and stability of 2D transistors.[12, 13] Nowadays, a variety of approaches have been experimentally and computationally demonstrated to modulate the thermal transport properties of 2D materials, such as twist angle of adjacent layers,[14-17] electrical regulation,[18] isotope-engineering,[19, 20] atomic vacancies,[21-23] and in-plane and out-of-plane strain engineering.[22, 24-27] Among them, twist engineering is a reversible process that does not destroy the internal structure of the heat-transporting material, and this physical modification method has good feasibility. Recent research results show that both the in-plane and cross-plane thermal conductivities of twisted bilayer van der Waals (vdW) heterostructures can be effectively regulated by their interfacial twist angle.[28-34] Based on homogeneous twisted hexagonal 2D materials, Ouyang et al.[35] revealed the underlying mechanism as the angle dependence of phonon−phonon couplings across the twisted interfaces, similar behavior has been observed for twisted graphene/*h*-BN interfaces.[36] Due to the structural anisotropy, the 2D layered materials exhibit high in-plane thermal conductivity but very low cross-plane thermal conductivity, and their ratio is defined as the thermal conductivity anisotropy ratio. Recently, a world record thermal conductivity anisotropy ratio in the randomly twisted $MoS_2$ has been experimentally reported by Kim et al.[37] The reason is that the twisted engineering significantly attenuates the out-of-plane thermal conductivity of $MoS_2$, while the change in the in-plane thermal conductivity is minimal.[19, 38, 39] This feature, however, cannot be captured by recent molecular dynamics (MD) simulations.[17, 40]

In this work, non-equilibrium molecular dynamics (NEMD) simulations have been carried out to gain a comprehensive understanding of the thermal transport properties of $MoS_2$. The cross-plane and in-plane thermal conductivity of $MoS_2$ has been investigated with various twisted angles using a recently developed registry-dependent interlayer potential (ILP)[41] that has been proved to capture well both the frictional properties and phonon spectra of layered materials.[5, 7, 35] Our results demonstrate that the twist angle significantly reduces the cross-plane thermal conductivity of 12-layered $MoS_2$ from 0.094 $Wm^{-1}K^{-1}$ to about 0.037 ~ 0.075 $Wm^{-1}K^{-1}$, which is in good agreement with the experimental value (0.057 ± 0.003 $Wm^{-1}K^{-1}$).[37] Furthermore, the computational results indicate that the in-plane thermal conductivity of twisted $MoS_2$ is insensitive to the interfacial twist angle, which is consistent with the experimental observations. To explain the mechanism behind this phenomenon and quantify the phonon contribution in the frequency domain, the spectral heat current (SHC) decomposition method has been employed to capture the insights of deterioration of the cross-plane thermal conductivity under different interfacial twist angles. Finally, the effect of twisted angles on the thermal conductivity anisotropy ratio in homogeneous $MoS_2$ has been discussed, emphasizing that twist engineering is an effective tool for modulating the thermal transport properties of 2D layered materials.

**Models**

The model system consists of two AA′-stacked $MoS_2$ slabs with the same number of layers and there is an interfacial twist angle $\theta$ between them, leading to the formation of moiré superlattices, as is shown in **Figure 1a**. To get reliable results, the edge and strain effects have been carefully eliminated by constructing the periodic twisted $MoS_2$ system (see *Methods*). **Figure 1b** and **c** illustrate the NEMD setup for calculating the cross-plane and in-plane thermal conductivity of the twisted MoS$_2$, respectively (see *Methods* for details). All MD simulations are performed with the LAMMPS package.[42]

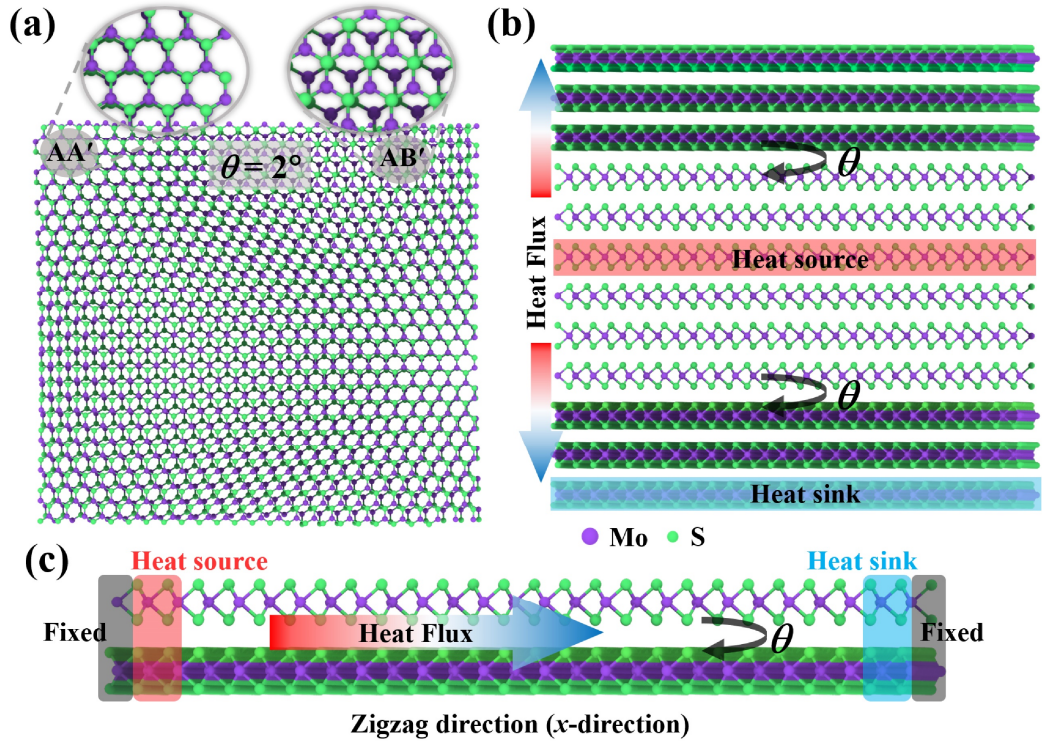

**Figure 1.** The schematic model of twisted MoS$_2$ for calculating the cross-plane and in-plane thermal conductivities. (a) Top view of bilayer MoS$_2$ with a twist angle of $\theta$ = 2°, which contains both AA′- and AB′-stacked domains. (b) Multilayer twisted MoS$_2$ model for calculating the cross-plane thermal conductivity. A thermal bias is induced by applying Langevin thermostats to the two layers marked by dashed red (Heat source) and blue (Heat sink) rectangles. The arrows indicate the direction of the vertical heat flux. Periodic boundary conditions are applied in the vertical direction, resulting in two twisted interfaces, and heat flows are in opposite directions across these interfaces. (c) Bilayer MoS$_2$ structure with a twisted angle $\theta$ for calculating the in-plane thermal conductivity. The atomic velocities and forces at both ends of the model are set to zero to simulate adiabatic walls.

To model the thermal transport behavior of MoS$_2$, we employed the REBO potential[43] and the state-of-the-art anisotropic interlayer ILP[41] to describe the intralayer interactions within each MoS$_2$ layer and the interlayer interaction between MoS$_2$ layers, respectively. The phonon spectra of bulk MoS$_2$ predicted by these force fields show good agreement with the experimental data, while the combination of REBO potential and the Lennard-Jones (LJ) potential considerably underestimates the out-of-plane phonon energies compared to experimental data (see **Figure S1**), indicating ILP is more accurate for the description of thermal transport properties of layered materials.[35] To check the effect of intralayer potential on the thermal conductivity of twisted MoS$_2$, the Stillinger-Weber (SW)[44] potential is adopted in combination with the ILP for the calculation of the cross-plane thermal conductivity of twisted MoS$_2$ and similar twist-angle dependence is observed (see **Figure S2**).

**Results and Discussions**

**Cross-plane thermal conductivity of twisted MoS$_2$**

We first investigated the effect of twist angle on the thermal transport properties across the twisted MoS$_2$ stacks. **Figure 2a** presents pronounced dependences of the cross-plane thermal conductivity ($\kappa_{\mathrm{CP}}$) of the entire stacks on the interfacial twist angle for both model systems consisting of 8 (red circle) and 12 (black triangle) layers of MoS$_2$. The $\kappa_{\mathrm{CP}}$ for both systems decreases quickly in the range of 0 ~ 5° and saturates above 5°, with an overall 2-3 fold reduction. Similar results have been observed for multilayer twisted graphene and $h$-BN systems.[35] Notably, the experimental value of $\kappa_{\mathrm{CP}}$ (0.057 ± 0.003 Wm$^{-1}$K$^{-1}$)[37] for randomly twisted MoS$_2$ lies well between the results of our simulations (see the blue line in **Figure 2a**), which indicates the accuracy of the anisotropic ILP for mimicking the vdW interaction across the multilayer MoS$_2$ stacks. Besides the twist-angle dependence, there is a tendency for $\kappa_{\mathrm{CP}}$ to weaken as the thickness decreases. This size effect obtained in the NEMD simulations originates from the large mean free path (MFP) of low-frequency phonons across the layers of the 2D layered materials,[38, 45, 46] which will be discussed later in detail.

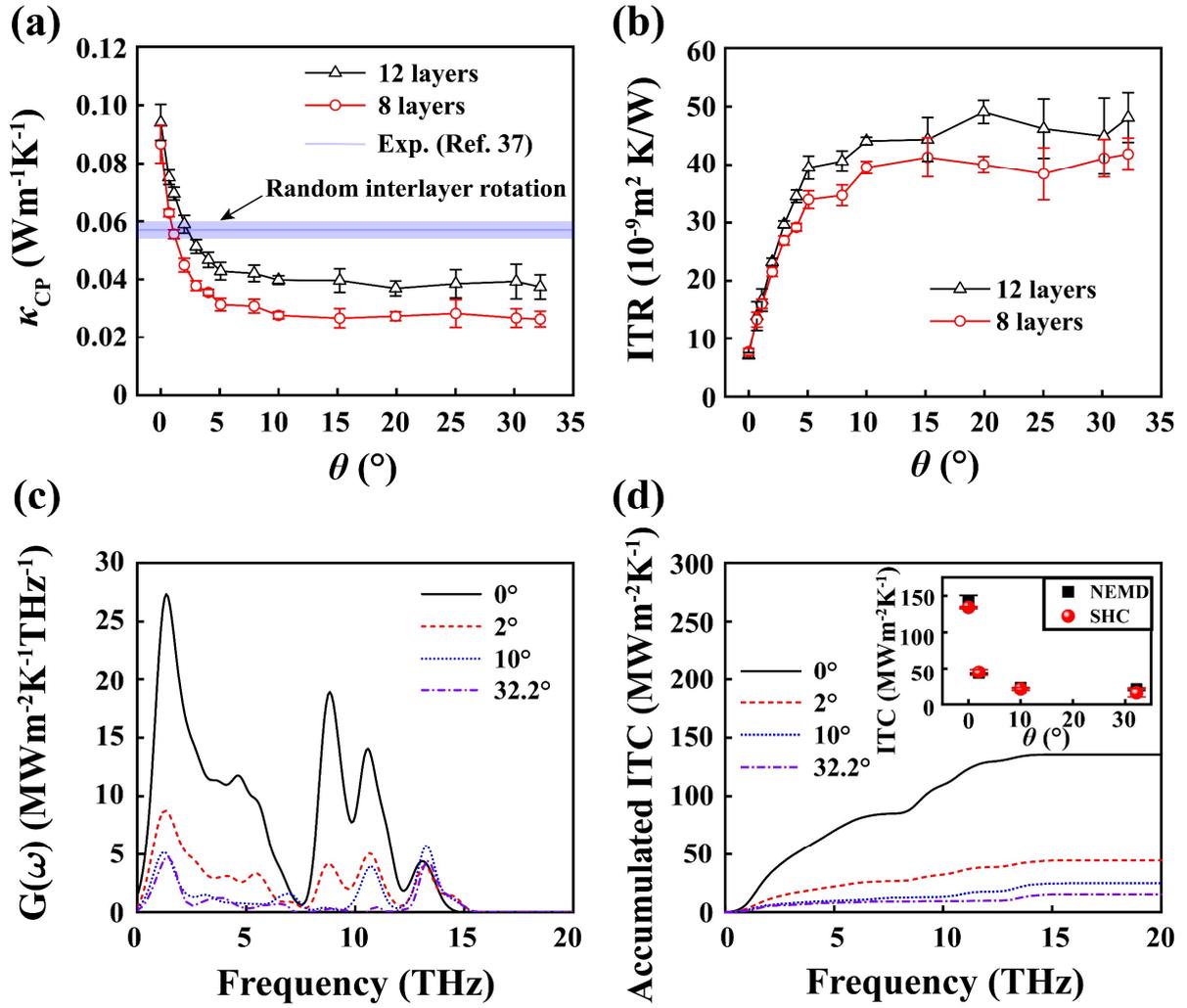

**Figure 2.** Twist-angle dependence of cross-plane thermal conductivity ($\kappa_{CP}$) (a) of the entire stacks and the interfacial thermal resistance (b) of the twisted contact formed between the optimally stacked slabs of MoS$_2$. The solid blue line in panel (a) marks the experimental value of $\kappa_{CP}$ for multilayer MoS$_2$ with randomly twisted interfaces obtained from time domain thermoreflectance measurements.[37] (c) Spectral phonon thermal conductance $G(\omega)$ across the twisted interface (see **Figure 1b**) that is calculated by the SHC method for various twisted angles. (d) The accumulated interfacial thermal conductance comes from the $G(\omega)$ integral, and the inset shows the comparison between SHC calculations and NEMD simulations.

To achieve a more comprehensive understanding of the impact that interfacial twist angles have on $\kappa_{CP}$, the interfacial thermal resistance (ITR), which measures the heat transport resistance of the two adjacent layers forming the twisted interface (see **Figure 1b**), is investigated. From **Figure 2b**, we can conclude that the ITR of the twisted interface shows a weak dependence on the sample thickness but strong dependence on the interfacial twist angle, with an overall 6-fold increase from 0 to 10°, such behavior is similar to that observed in hexagonal 2D materials,[35] indicating that the interfacial twist angle is an important factor for regulating and understanding the thermal transport properties in layered materials.

To further explore the microscopic mechanisms of the strong twist-angle dependence of $\kappa_{CP}$, we carry out the SHC decomposition analysis for the 12-layered twisted MoS$_2$ stacks (see *Methods* for details). More specifically, the spectral phonon thermal conductance $G(\omega)$ is calculated for four interfacial twist angles ($\theta$ = 0°, 2°, 10°, and 32.2°). **Figure 2c** and **d** present that the contribution to the $\kappa_{CP}$ of MoS$_2$ comes from relatively low-frequency phonons (<15 THz), which have longer wavelengths and stronger phonon coupling relative to that of high-frequency phonons, thus dominating the cross-plane thermal transport for layered MoS$_2$ materials.[36, 47, 48] In addition, the peak of $G(\omega)$ in the 0–5 THz and 7.5–12.5 THz domains decreases significantly as the interfacial twist angle increases from 0° to 32.2°, indicating that the twist engineering mainly weakens the thermal transport capability of phonons at these two specific frequency regions. Moreover, we calculated the interfacial thermal conductance (ITC) of the twisted interface using the SHC method (as shown in **Figure 2d**) and compared it to NEMD results (here the ITC is defined as ITC ≡ 1/ITR). As seen from the inset in **Figure 2d**, the ITC calculated by the SHC method is in excellent agreement with NEMD simulation results for all studied twist angles, indicating the reliability of the SHC method for capturing the characteristics of phonons on the spectrum.[49, 50]

To better align our calculations with the experimental values, we computed the cross-plane thermal conductivity of MoS$_2$ systems ($\theta$ = 0°) with varying stack thicknesses (8-56 layers). As is shown in **Figure 3a**, the cross-plane thermal conductivity increases monotonically with increasing thickness and shows a tendency to converge for the larger number of layers. This convergence characteristic is also observed for the interfacial thermal resistance (**Figure 3b**). We used the linear extrapolation method to calculate the cross-plane thermal conductivity of bulk MoS$_2$, which is about ~0.24 Wm$^{-1}$K$^{-1}$, close to the experimentally reported value (~0.27 Wm$^{-1}$K$^{-1}$) for bulk MoS$_2$ containing vertical grains.[51] However, for the pristine bulk MoS$_2$, the reported values of $\kappa_{CP}$ are in the range of 2–5 Wm$^{-1}$K$^{-1}$,[38, 51-53] which were substantially larger than our calculated value. This discrepancy is attributed to the fact that the thickness of the MoS$_2$ sample (up to 56 layers) is far below its cross-plane MFP (>200 nm).[38] However, the size of twisted MoS$_2$ stacks is larger than its cross-plane MFP (~2 nm),[37] thus our NEMD simulations for twisted MoS$_2$ stacks show good agreement with the experimental values (see **Figure 2a**).

Next, we calculate the spectral phonon thermal conductance $G(\omega)$ of untwisted MoS$_2$ by varying the layer number from 8 to 56. The thickness variation mainly affects the phonons at low frequencies and is particularly sensitive to the peak at about 1 THz, as shown in **Figure 3c.** This phenomenon occurs when low-frequency phonons with long wavelengths are allowed in samples of greater thickness and dominate the thermal conductivity across the plane. The accumulated ITC of different layers, as shown in **Figure 3d**, also increases with the number of layers, consistent with the layer-dependent cross-plane thermal conductivity of MoS$_2$. In addition, comparing the ITC calculated by the SHC method with the NEMD simulation, we find that the ITCs of the corresponding layers are essentially close and demonstrate similar layer-dependent properties in the inset of **Figure 3d.** Thus, the $G(\omega)$ for different layers helped us determine the contribution of low-frequency (long-wavelength) phonons (0-5 THz) as the dominant responsible for increasing the out-of-plane thermal conductivity of MoS$_2$ with the increasing number of layers.

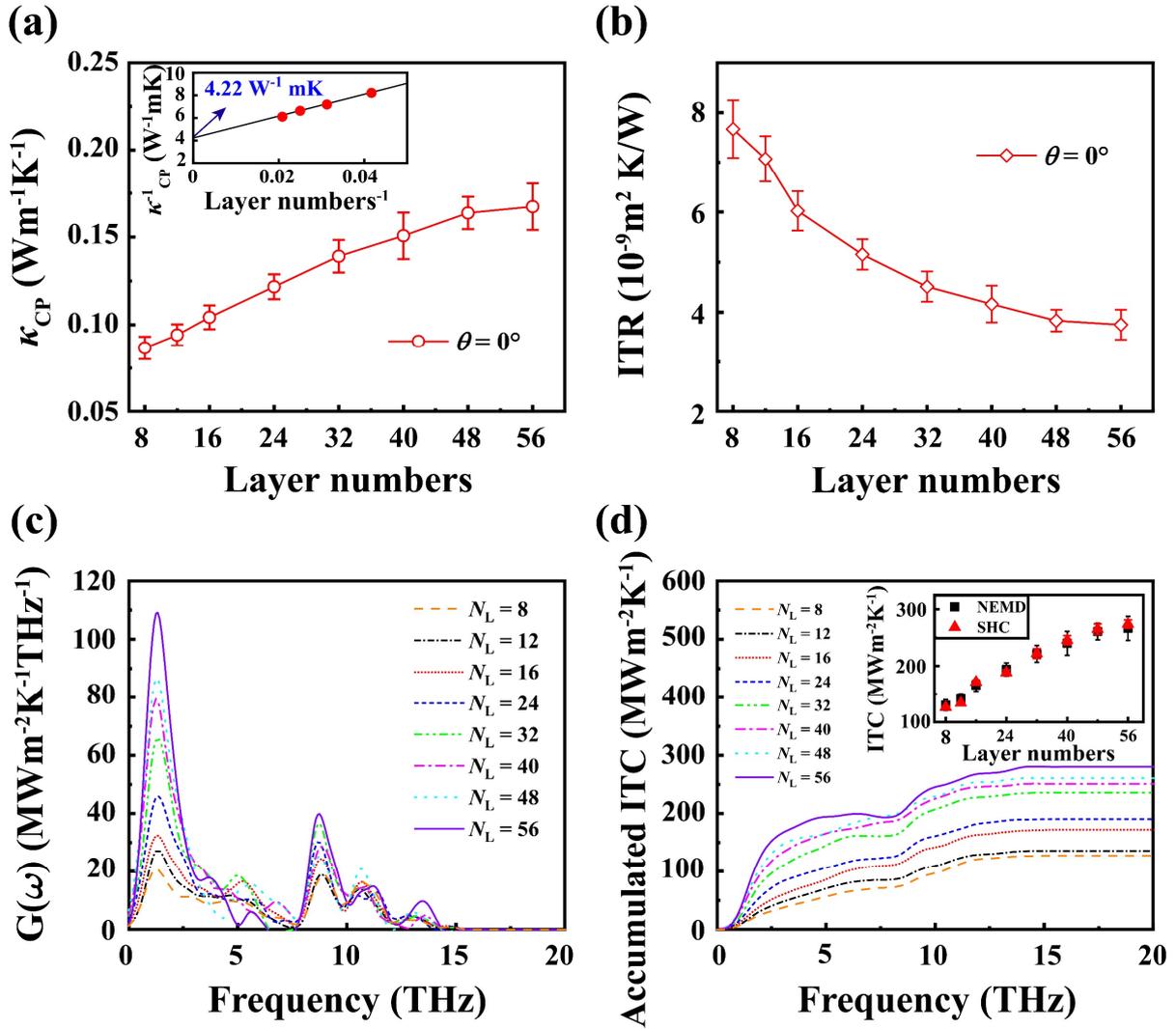

**Figure 3.** Thickness dependence of the (a) cross-plane thermal conductivity and (b) interfacial thermal resistance for untwisted MoS$_2$ stacks ($\theta = 0°$). Panels (c) and (d) show the frequency-dependent spectral phonon thermal conductance $G(\omega)$ and ITC with various thicknesses of the untwisted MoS$_2$ stacks, respectively. The inset in panel (a) shows the results using the linear extrapolation method. The inset in panel (d) compares the results of SHC calculations to NEMD simulations. $N_L$ represents the number of layers in the homogenous MoS$_2$ stacks.

**In-plane thermal conductivity of twisted MoS$_2$**

We calculated the in-plane thermal conductivity ($\kappa_{\text{IP}}$) of twisted MoS$_2$ with the heat current flowing along the *x*-direction (zigzag) of the MoS$_2$ lattice for a series of interfacial twist angles and sample sizes (see **Figure 4a-b**). The width of MoS$_2$ that is perpendicular to the direction of heat current (i.e., armchair direction) is set as 13 nm, which is large enough to get converged $\kappa_{\text{IP}}$ of MoS$_2$ (see **Figure S3**). From **Figure 4a** and **b**, we find that $\kappa_{\text{IP}}$ show a weak and strong dependence on the twist angle and sample size, respectively. The former is consistent with recent experimental measurements for randomly stacked MoS$_2$,[37] and NEMD simulations with LJ potential for bilayer MoS$_2$/MoSe$_2$ heterostructures.[54] While in the latter case, $\kappa_{\text{IP}}$ varies from 2.3 Wm$^{-1}$K$^{-1}$ to 12.8 Wm$^{-1}$K$^{-1}$ as the length increases from 5 nm to 100 nm. To clarify the weak twist-angle and strong size dependences of the in-plane thermal conductivity of MoS$_2$, we calculated the in-plane spectral phonon thermal conductance $G(\omega)$ and the corresponding accumulated $\kappa_{\text{IP}}$ using the SHC method for four interfacial angles (**Figure 4c**) and six sample sizes (**Figure 4d**). The calculated $G(\omega)$ is almost twist-angle independent but strongly depends on the sample size, with the main contribution coming from phonons in the range of 0-8 THz.

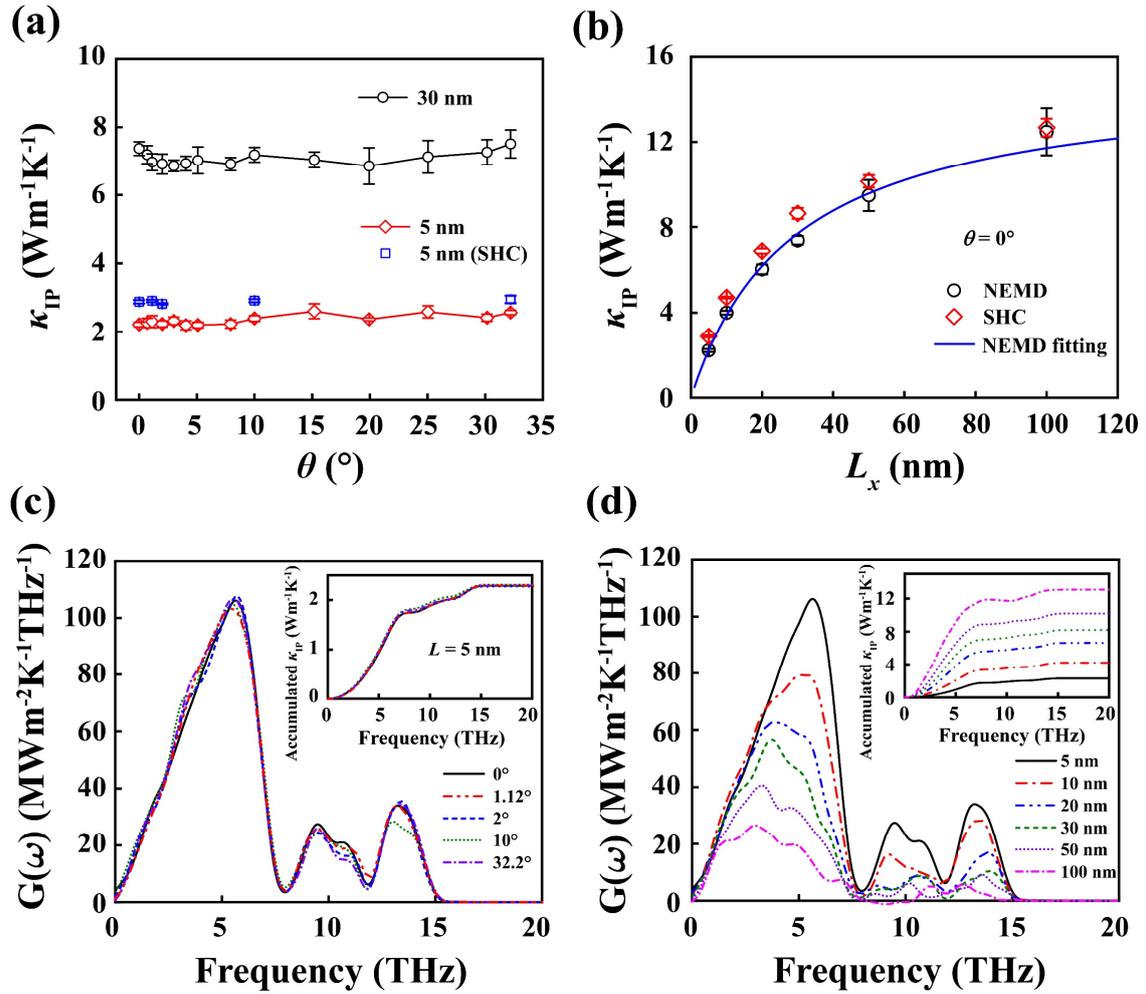

**Figure 4.** The in-plane thermal conductivity ($\kappa_{IP}$) of bilayer MoS$_2$ as a function of the twist angle (a) and the length along the direction of the heat flux (b). The solid blue line in panel (b) is the extrapolated fit of NEMD simulations using Eq. (9). Panels (c) and (d) are the corresponding spectral phonon thermal conductance $G(\omega)$ with respect to the twist angle and length, respectively.

As illustrated in **Figure 4d,** the $G(\omega)$ for phonon frequency larger than 1 THz shows a strong dependence on the sample length, which is attributed to the anharmonic effect.[55-57] The weak size dependence of $G(\omega)$ for the low-frequency phonons (<1 THz) indicates a ballistic phonon transport in this region. The inset of **Figure 4d** demonstrates the size dependence of calculated accumulated in-plane thermal conductivity $\kappa_{IP}$ and gives consistent results with that obtained by NEMD simulations. To evaluate the $\kappa_{IP}$ of MoS$_2$ in the bulk limit, we adopted the following straightforward extrapolation equation:[58]

$$\frac{1}{\kappa(L_x)} = \frac{1}{\kappa_0}\left(1 + \frac{\lambda}{L_x}\right) \tag{1}$$

where $\kappa_\infty = \kappa(L_x \to \infty)$ is the extrapolated $\kappa_{\text{IP}}$ in the infinite-length limit, and $\lambda$ is an effective MFP. **Figure S5** shows that a linear fit to the NEMD data gives $\kappa_\infty = 15.1 \pm 1.2$ Wm$^{-1}$K$^{-1}$ and $\lambda = 29.0 \pm 3.0$ nm. While for recent experimental measurements, the fitted value is $\kappa_\infty = 44 \pm 6$ Wm$^{-1}$K$^{-1}$.[37] In our NEMD simulations, the lengths along the heat flux direction are varied from 5 nm to 100 nm, smaller to the largest MFP of the phonons of MoS$_2$ that dominate the thermal transport,[54, 59, 60] the extrapolated $\kappa_\infty$ of the bilayer MoS$_2$ is thus lower than the experimentally obtained value. However, the fact that both values of $\kappa_\infty$ are in the same order indicates the MFP of MoS$_2$ is in the order of a few tens of nanometers, which is reachable in NEMD simulations. Therefore, the $\kappa_{\text{IP}}$ of MoS$_2$ calculated from the NEMD simulations is qualitatively consistent with the experimental observations.

**Thermal conductivity anisotropy ratio of twisted MoS$_2$**

Finally, we discuss the effect of interlayer twisted angle on the thermal conductivity anisotropy ratio along the cross-plane ($\kappa_{CP}$) and in-plane ($\kappa_{IP}$) directions. The cross-plane anisotropy ratio is defined as $\rho_\perp = \kappa_{IP}/\kappa_{CP}$ and the in-plane anisotropy ratio is defined as $\rho_\parallel = \kappa_{IP}^{max}/\kappa_{IP}^{min}$, where $\kappa_{IP}^{max}$ and $\kappa_{IP}^{min}$ are the maximum and minimum in-plane thermal conductivities, respectively. Before calculating the anisotropy, we compare the reported results for the in-plane thermal conductivity of bilayer twisted MoS$_2$, as shown in **Figure 5a**. The NEMD simulation results of Mandal *et al.*[40] using the Müller-Plathe method show that the $\kappa_{IP}$ of MoS$_2$ first reduces and then increases with increasing twist angle, while Nie *et al.*[17] present the "*w*-shaped" twist-angle-dependent $\kappa_{IP}$ using the LJ potential. Here, our NEMD simulations based on ILP demonstrate a weak twist-angle dependence of in-plane thermal conductivity, which is consistent with the recent experimental data (see **Figure 5a**). The above results indicate that the predicted thermal transport properties of layered materials are sensitive to the simulation protocol and the adopted interlayer force field, thus careful benchmark tests should be performed to get reliable results.[35]

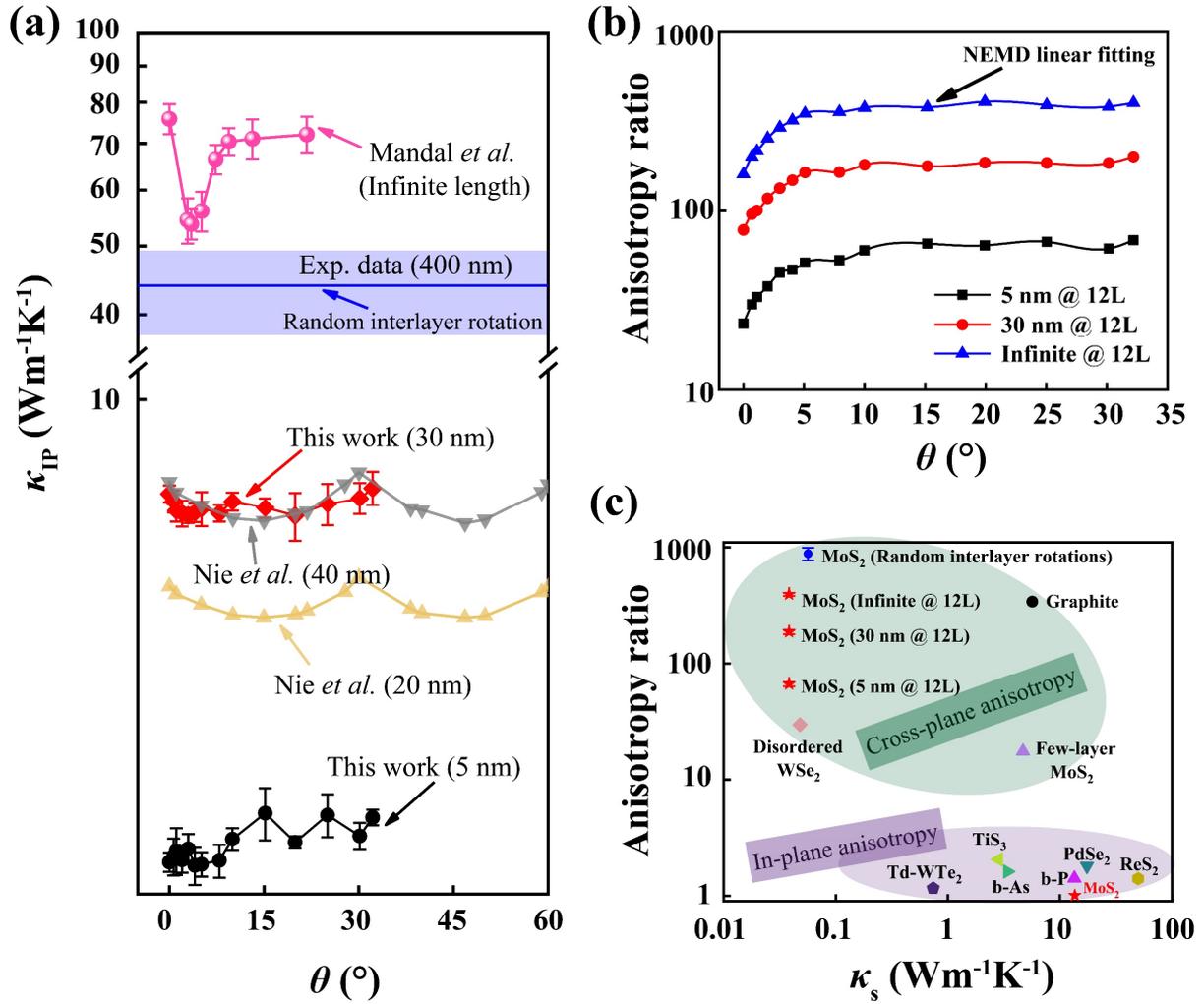

**Figure 5.** (a) Dependence of the in-plane thermal conductivity of bilayer MoS$_2$ on the twist angle. Simulation data is extracted from Fig. 8 of Ref. 17 and Fig. 2 of Ref. 40, and the experiment results are measured using time-domain thermoreflectance (TDTR) for multilayer MoS$_2$ with random interlayer rotation.[37] (b) The thermal conductivity anisotropy ratio calculated from our NEMD simulations for a 12-layer MoS$_2$ stack with various twisted angles. (c) Comparison of anisotropy ratio $\rho$ (y-axis), $\kappa_s$ (x-axis) for different anisotropic thermal conductors, where the $\kappa_s$ represents the thermal conductivity along the axis with the slowest heat current. The data for in-plane thermal conductivity anisotropy ratio includes Td-WTe$_2$,[61] titanium trisulfide (TiS$_3$),[62] black arsenic (b-As),[63] black phosphorus (b-P),[64] pentagonal PdSe$_2$,[65] rhenium disulfide (ReS$_2$),[66] and twisted MoS$_2$ (our results, see **Figure S6**). The data for the thermal conductivity anisotropy ratio includes randomly twisted MoS$_2$,[37] disordered

WSe$_2$,[67] few-layer MoS$_2$,[53] and graphite.[68]

**Figure 5b** presents the calculated cross-plane thermal conductivity anisotropy ratios of twisted MoS$_2$ using NEMD simulations with various sizes. The results show obvious twist-angle and size dependences (~3-fold strengthen), which is resulted from the twist-angle dependence of $\kappa_{\mathrm{CP}}$ and size dependence of $\kappa_{\mathrm{IP}}$, respectively. In the latter case, we considered the anisotropy ratio obtained from the extrapolated $\kappa_{\mathrm{IP}}$ in the infinite-length limit (see blue triangles in **Figure 5b**). It clearly shows that twist engineering can effectively tune the anisotropy ratio of MoS$_2$. In addition, we compared the anisotropy ratio $\rho$ of the twisted MoS$_2$ to that of the reported values for various thermal conductors from the literature. As shown in **Figure 5c**, the in-plane anisotropy ratio of twisted MoS$_2$ is nearly 1 according to our simulations, which is consistent with experimental measurements for various thermal conductors (barely exceeding 2). While for the cross-plane thermal conductivity anisotropy ratio, we report a value of up to 400 for twisted MoS$_2$ (corresponding to "Infinite@12L" in **Figure 5c**), which is in the same order as the world record (~900).[37] Comparing to disordered layered WSe$_2$ ($\rho \approx 30$)[67] and bulk MoS$_2$ ($\rho \approx 20$),[53] the twisted MoS$_2$ has a larger $\rho$ because interlayer rotation reduces cross-plane thermal conductivity significantly but has a minor effect on the excellent in-plane heat transporting properties. Our results show that the twist MoS$_2$ stacks can be used as an excellent thermal conductor or thermal insulator by tuning the direction of heat current through it.

## Conclusion

In conclusion, we systematically investigated both the cross-plane and in-plane thermal transport properties of homogeneous twisted MoS$_2$ vdW structures based on the NEMD simulations with the state-of-art force fields. While the cross-plane thermal conductivity presents a pronounced twist angle dependence, the in-plane thermal conductivity is insensitive to it. Both the cross-plane and in-plane thermal conductivities show a strong size effect, with a bulk limit of ~0.24 and ~15.1 Wm$^{-1}$K$^{-1}$, respectively. The former value is close to the recent experimental measurements (~0.27 Wm$^{-1}$K$^{-1}$), which indicates the accuracy of the interlayer force field developed for MoS$_2$. The great difference between the in-plane and cross-plane thermal conductivities (with a ratio up to 400) shows the highly anisotropic thermal transport properties of twisted MoS$_2$ vdW structures, which also presents a pronounced twist-angle dependence. We thus expect twist engineering to be an effective and generalizable way to reduce cross-plane thermal conductivity and potentially tune anisotropic thermal properties in various 2D materials.

## Methods

**Building commensurate structures for twisted MoS$_2$.** For bilayer twisted MoS$_2$, the superlattice is strictly periodic only at certain special angles, since the moiré interference pattern does not exactly match the general lattice period. Here, we defined the primitive lattice vector of the bottom MoS$_2$ layer as $\boldsymbol{a}_1 = a(1,0)$ and $\boldsymbol{a}_2 = a(1/2, \sqrt{3}/2)$, where $a$ is the lattice constant of monolayer MoS$_2$. Then the lattice vector of the bottom MoS$_2$ layer and top MoS$_2$ layer can be given by $\boldsymbol{L}_1 = i_1\boldsymbol{a}_1 + i_2\boldsymbol{a}_2$ and $\boldsymbol{L}_2 = j_1\boldsymbol{a}_1 + j_2\boldsymbol{a}_2$, respectively, where $\boldsymbol{L}_1$ is equals to $\boldsymbol{L}_2$ with certain integers $i_1$, $i_2$ and $j_1$, $j_2$. Thus, the exact superlattice period is then given by:[69]

$$L = |i_1\boldsymbol{a}_1 + i_2\boldsymbol{a}_2| = a\sqrt{i_1^2 + i_2^2 + i_1 i_2} = \frac{|i_1 - i_2|a}{2\sin(\theta/2)} \qquad (2)$$

Where the interfacial twist angle $\theta$ is equal to the angle between two lattice vectors $\boldsymbol{L_1}$ and $\boldsymbol{L_2}$. When building commensurate structures for twisted MoS$_2$, we always rotated the supercell such that its lattice vector is $\boldsymbol{L_1} = L(1,0)$ and $\boldsymbol{L_2} = L(1/2, \sqrt{3}/2)$. The parameters used to construct periodic supercells of various interfacial twist angles can be extracted from Ref. 35.

**NEMD simulations for calculating the thermal conductivity.** To calculate the cross-plane thermal conductivity of twisted MoS$_2$ (see **Figure 1b**), periodic boundary conditions are used in all directions, and the time step for propagating the equations of motion is set to 0.5 fs. A Nosé–Hoover thermostat[70, 71] with a time constant of 0.25 ps is used to maintain a stable temperature. We first equilibrated the system in the NPT ensemble at a temperature of $T$ = 300 K and zero pressure for 500 ps to relax the box for minimizing the effect of residual stress (see **Figure S4**). After the thermal equilibration, the heat source ($T$ = 375 K) and heat sink ($T$ = 225 K) are set at the bottom and middle layers to establish a temperature gradient, which is used by the Langevin thermostat method.[72, 73] Then the system is switched to NVE ensemble for 750 ps to equilibrate. The cross-plane thermal conductivity of twisted MoS$_2$ is calculated by the MD data from the last 500 ps, and ten different data sets calculated over a time interval of 50 ps are used to estimate the statistical errors. **Figure 1c** shows the bilayer MoS$_2$ model used to investigate the in-plane thermal conductivity. The shrink-wrapped boundary conditions are applied in the $x$- (zigzag) and $z$-direction, and the periodic boundary conditions are assigned to the $y$-direction (armchair). The bilayer MoS$_2$ is divided into the $N$ slabs along the $x$-direction, and two slabs at both ends are fixed to simulate adiabatic walls.

For the sake of finding the steady-state system, the models are equilibrated in a Nosé–Hoover thermostat[70, 71] for 300 ps at 300K. Then the system is coupled with an NVT ensemble at the heat source and sink regions adjacent to the fixed slabs with 350 K and 250 K to obtain the temperature distribution in other slabs and the NVE ensemble is applied in interior regions. The NEMD for the bilayer MoS$_2$ is performed for 2 ns. After obtaining the steady non-equilibrium temperature profile in the models, i.e., when the temperature gradient illustrates that the systems are already relaxed for a sufficient time and have reached steady (see **Figure S7**), the total thermal conductivity can then be calculated from one-dimensional Fourier's law in the direction of heat flux as follows:

$$\kappa_{\text{tot}} = \frac{\dot{Q}/A}{dT/dL} \qquad (3)$$

where $A$ is the cross-sectional area, $\dot{Q}$ is the calculated heat current, and $dT/dL$ is the effective temperature gradient along the heat flux direction. To calculate the cross-sectional area, the thickness of monolayer MoS$_2$ is assumed to be 0.621 nm. According to the definition of the Kapitza resistance,[74] the ITR can be calculated as:[35]

$$\left(\frac{N}{2}-3\right)R_{\text{AA}'} + R_\theta = \left[\left(\frac{N}{2}-3\right)d_{\text{AA}'} + d_\theta\right]/\kappa_{\text{CP}} \qquad (4)$$

where $d_\theta$ is the interlayer distance between two adjacent twisted layers of MoS$_2$ and $N$ is the number of layers of the entire system. For the AA′-stacked system ($\theta = 0°$), $R_{\text{AA}'}$ equals to $R_\theta$.

**Spectral heat current calculation.** To quantitatively clarify the role of the twist angle on phonon thermal transport, the SHC decomposition through the twisted interface is investigated. The interparticle spectral heat current is defined as:[55, 56]

$$q_{i \to j}(\omega) = -\frac{2}{\omega t_{\text{sim}}} \sum_{\alpha,\beta \in \{x,y,z\}} \text{Im} \langle \hat{v}_i^\alpha(\omega) * K_{ij}^{\alpha\beta} \hat{v}_j^\beta(\omega) \rangle, \qquad (5)$$

where $\omega$ is the phonon frequency and the $t_{\text{sim}}$ is the total simulation time. $\hat{v}_i^\alpha(\omega)$ and $\hat{v}_j^\beta(\omega)$ are the Fourier-transformed atomic velocities of atom $i$ in direction $\alpha$ and atom $j$ in direction $\beta$, respectively. The $K_{ij}^{\alpha\beta}$ represents the harmonic interatomic force constant matrix. The $\langle \cdot \rangle$ represents the ensemble average, which is assumed to be equal to the time average due to ergodicity. We calculate the force constant by the finite displacement method by using a Python code, which is publicly available.[75] The atom $i$ is moved to the $\pm x$, $\pm y$, and $\pm z$ directions with a small displacement value Δ = 0.01 Å (keeping all the other atoms at their relaxed positions), respectively. After each atom has been displaced in three directions, the force constant can be calculated as follows:

$$K_{ij}^{\alpha\beta} = \frac{F_j^{\beta-} - F_j^{\beta+}}{2\Delta} \tag{6}$$

Where $F_j^{\beta-}$ and $F_j^{\beta+}$ denote the forces in the $\beta$-direction of atom $j$ when $i$ atoms are displaced to the $-\alpha$ and $+\alpha$ directions, respectively. The spectral phonon thermal conductance $G(\omega)$ can be obtained by summing over all pairs of atoms in a twisted interface divided by the temperature difference ($\Delta T$) and cross-sectional area ($A$):

$$G(\omega) = \frac{1}{A\Delta T}\sum_{i\in L}\sum_{j\in R} q_{i\to j}(\omega) \tag{7}$$

Two adjacent regions in the middle of the model are selected for the "$L$" and "$R$" groups. From the spectral phonon thermal conductance $G(\omega)$, we also calculate the in-plane spectrally decomposed thermal conductivity:

$$\kappa(\omega) = G(\omega) \cdot L \tag{8}$$

Accordingly, by summing the $\kappa(\omega)$ at different frequencies, we calculate the thermal conductivity $\kappa$ of the corresponding length $L$ from the SHC method:

$$\kappa_{\text{tot}} = \int_0^\infty \frac{d\omega}{2\pi}\kappa(\omega) \tag{9}$$

**Acknowledgments**

The authors acknowledge support from National Natural Science Foundation of China (Nos. 12102307), the Natural Science Foundation of Hubei Province (2021CFB138), the start-up fund of Wuhan University, the National Key R&D Project from Ministry of Science and Technology of China (Grant No. 2022YFA1203100) and the Research Grants Council of Hong Kong (Grant No. AoE/P-701/20).

## *Supporting Information for*

# Twist-Dependent Anisotropic Thermal Conductivity in Homogeneous MoS$_2$ Stacks

Wenwu Jiang[1#], Ting Liang[2#], Jianbin Xu[2], Wengen Ouyang[1*]

[1]Department of Engineering Mechanics, School of Civil Engineering, Wuhan University, Wuhan, Hubei 430072, China

[2]Department of Electronic Engineering and Materials Science and Technology Research Center, The Chinese University of Hong Kong, Shatin, N.T., Hong Kong SAR, 999077, P. R. China

#: These authors contributed equally.
*E-mail: w.g.ouyang@whu.edu.cn (W. Ouyang)

This supporting information document includes the following sections:

1. Comparison of phonon spectrum calculated using the ILP and the Lennard-Jones potential
2. Effect of intralayer potential on the $\kappa_{\text{CP}}$ of twisted MoS$_2$ stacks
3. Converge tests
4. Extrapolation of the $\kappa_{\text{IP}}$ in the bulk limit from NEMD simulation
5. Anisotropic in-plane thermal conductivity of bilayer MoS$_2$
6. Temperature profiles

# 1. Comparison of the phonon spectrum calculated using the ILP and the Lennard-Jones potential

A proper description of the vdW interaction can reproduce the measured low-frequency phonon dispersion of bulk $MoS_2$ more accurately, which is very important for investigating the cross-plane thermal transport properties. To compare the accuracy of the two different interlayer force filed (ILP[1] and the Lennard-Jones (LJ)[2] potential), we calculated the phonon spectra for untwisted bulk $MoS_2$ using the REBO+ILP and REBO+LJ force fields, respectively. Here, the LJ parameters ($\sigma$ and $\varepsilon$) for Mo-Mo, Mo-S, and S-S are taken from the universal force field (UFF),[2] which are listed in **Table S1**, and could describe the interlayer coupling between two $MoS_2$ layers. **Figure S1a** shows the results of calculated phonon spectrum curves, however, the interlayer potential mainly affects the dispersion of low-energy out-of-plane branched (see **Figure S1b**). We find that the LJ potential considerably underestimates the out-of-plane phonon energies compared to that of ILP and the phonon spectrum calculated by REBO+ILP agrees well with the experimental data. This section supports the reliability of the ILP for calculating cross-plane thermal conductivity.

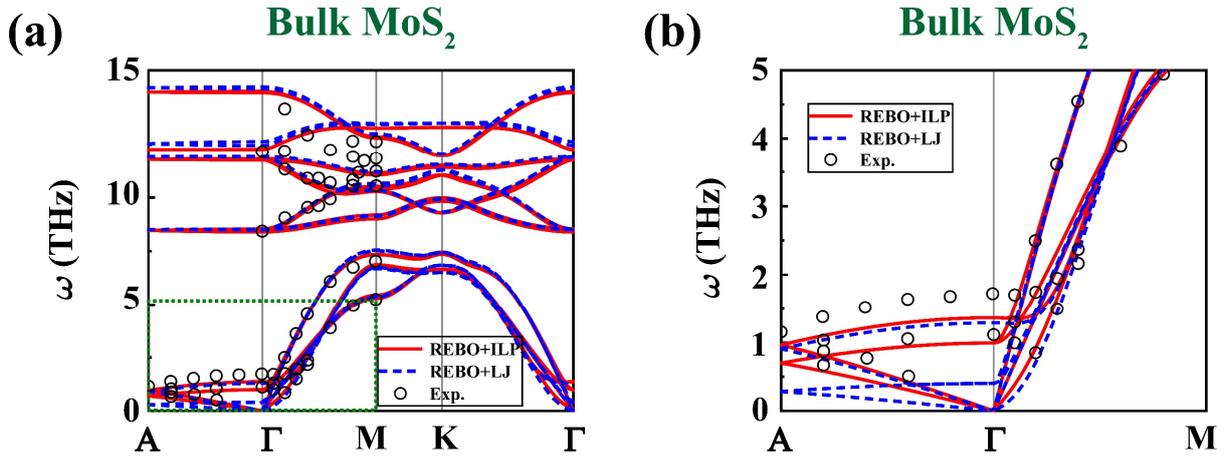

**Figure S1.** (a) Phonon spectrum of bulk $MoS_2$. Red solid lines and bulk dashed lines are dispersion curves calculated using the REBO + ILP and REBO + LJ force fields, respectively. Black circles present the experimental results of bulk $MoS_2$.[3] A zoom-in region (green dashed rectangle in panel (a)) on the low energy phonon modes around the $\Gamma$-point for bulk $MoS_2$ is shown in panel (b).

**Table S1.** Lennard-Jones parameters for MoS$_2$ interlayer interactions. The LJ parameters are taken from Ref. 2.

|  | Mo-Mo | Mo-S | S-S |
|---|---|---|---|
| σ (Å) | 2.719 | 3.157 | 3.595 |
| ε (meV) | 2.428 | 5.371 | 1.188 |

## 2. Effect intralayer potential on the $\kappa_{CP}$ of twisted MoS₂ stacks

A proper description of the intralayer interaction is essential for investigating the thermal transport properties. In the main text, we only show the cross-plane thermal conductivity ($\kappa_{CP}$) results of MoS₂ using the REBO[4] potential. However, SW[5] potential is also widely used to describe the intralayer interaction for the MoS₂ system. To investigate the effect of intralayer potential on the calculation of $\kappa_{cp}$, we calculated the thickness dependence and twist-angle dependence of $\kappa_{cp}$ using both the REBO and SW potentials, while keeping the ILP the same. The results are illustrated in **Figure S2**. **Figure S2a** presents that the $\kappa_{cp}$ of MoS₂ systems ($\theta = 0°$) obtained using both the REBO and SW potential increases monotonically with increasing thickness and shows a tendency to converge. The fitted $\kappa_{cp}$ for infinite thickness obtained by the linear extrapolation method is ~0.24 Wm⁻¹K⁻¹ and ~0.47 Wm⁻¹K⁻¹ for REBO and SW potentials, respectively. The calculated $\kappa_{cp}$ using SW potential exhibited a 2-fold increase compared to that calculated by the REBO potential. As the twist angle increases, the difference of $\kappa_{cp}$ for twisted MoS₂ stacks obtained using REBO and SW potential decreases dramatically, as shown in **Figure S2b**.

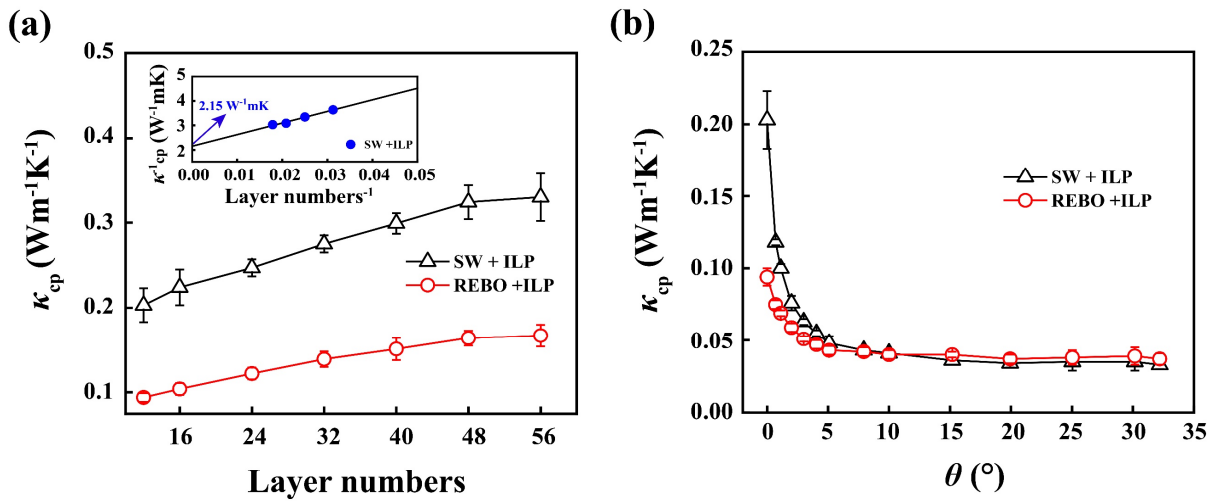

**Figure S2.** Thickness dependence (a) and twist-angle dependence (b) of the cross-plane thermal conductivity $\kappa_{cp}$ of the MoS₂ system. Red circles and black triangles represent results obtained using the REBO[4] potential and SW[5] potential, respectively.

## 3. Convergence tests

### 3.1 Convergence tests respect to the width of supercell

In this main text, the heat flow was applied in the *x*-direction (zigzag) when calculating the in-plane thermal conductivity of the untwisted $MoS_2$. To evaluate the convergence of our results for the lengths along the *y*-direction (armchair), we performed additional calculations for the aligned $MoS_2$ bilayer with $\theta = 0°$ (see **Figure S3**). Supercells of the armchair-direction lengths of 131.8 Å, 263.5 Å, 395.3 Å, and 527.1 Å are used for the aligned ($\theta = 0°$) interface, while the *x*-direction length is fixed to 50 Å. The corresponding calculated thermal conductivities are $2.217 \pm 0.069$ Wm$^{-1}$K$^{-1}$, $2.213 \pm 0.040$ Wm$^{-1}$K$^{-1}$, $2.203 \pm 0.034$ Wm$^{-1}$K$^{-1}$, and $2.249 \pm 0.040$ Wm$^{-1}$K$^{-1}$, respectively. These results show that the in-plane thermal conductivity we obtained for twisted $MoS_2$ is insensitive to the choice of supercell width (armchair direction) in the NEMD simulations.

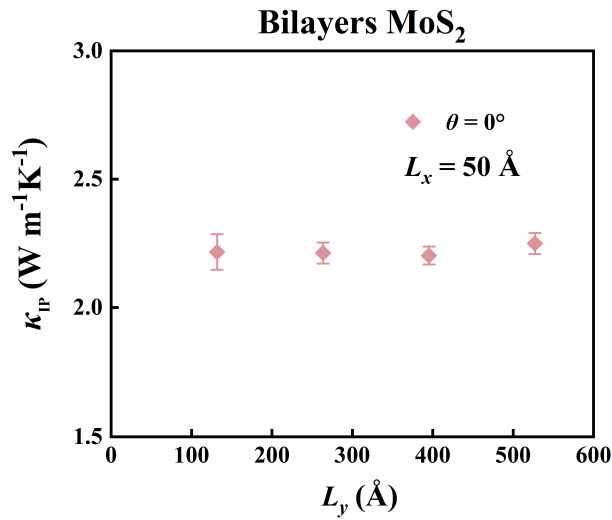

**Figure S3.** In-plane thermal conductivity of the aligned bilayer $MoS_2$ as a function of the size along the *y*-direction (armchair), while the heat current is applied along the *x*-direction (zigzag).

## 3.2 Convergence tests respect to the simulation time

To show the NEMD simulations for all model systems already reach the steady state, we present the time evolutions of the temperature of the thermostated layers in **Figure S4**, which clearly shows the steady-state heat current has been established. Here, we also found that the temperature difference between the thermostated layers increases with increasing the twist angle, which is attributed to the growing interfacial thermal resistance as the twist angle increases.[6]

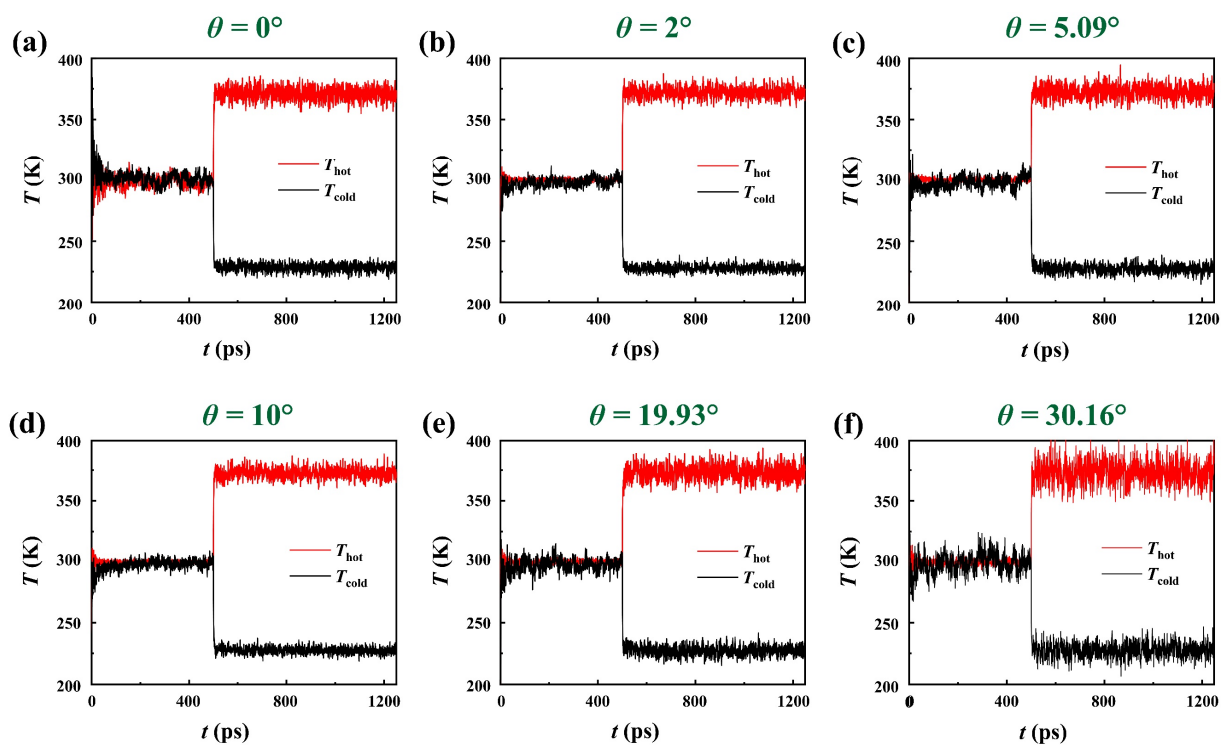

**Figure S4.** Time evolution of the temperature of the thermostated layers for 12 layers twisted $MoS_2$ with misfit angle (a) $\theta = 0°$, (b) $\theta = 2°$, (c) $\theta = 5.09°$, (d) $\theta = 10°$, (e) $\theta = 19.93°$, and (f) $\theta = 30.16°$.

## 4. Extrapolation of the $\kappa_{IP}$ in the bulk limit from NEMD simulations

It is well known that the strong size dependence of the thermal conductivity calculated by NEMD resulted from the much longer phonon mean free path compared to the sample size in the simulations. To get the size limit of the in-plane thermal conductivity of MoS2, we used the most straightforward extrapolation equation (see Eq. (1) in the main text). The fitting results of $1/\kappa$ and $1/L$ are shown in **Figure S5**, which gives $\kappa_\infty = 15.1 \pm 1.2 \ \text{Wm}^{-1}\text{K}^{-1}$ and $\lambda = 29.0 \pm 3.0$ nm.

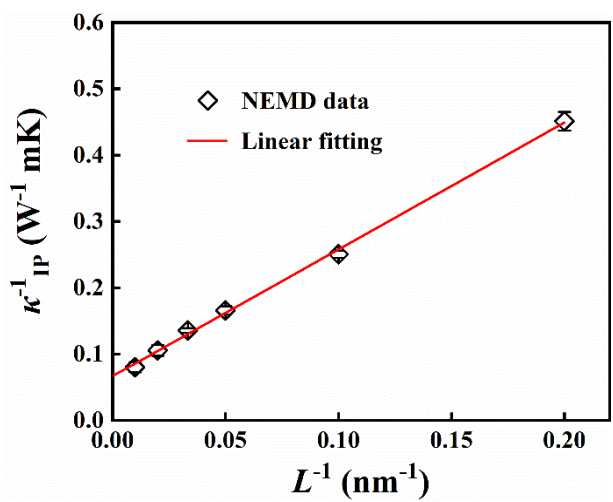

**Figure S5.** Inverse in-plane thermal conductivity $1/\kappa_{IP}$ as a function of the inverse simulated length $1/L$ in NEMD simulations. The in-plane thermal conductivity for infinitely long twisted MoS2 can be obtained by linear extrapolation at $1/L = 0$.

## 5. Anisotropic in-plane thermal conductivity of bilayer MoS$_2$

In this section, the in-plane thermal conductivity ($\kappa_{IP}$) along different directions is calculated to investigate the in-plane anisotropy ratio of bilayer MoS$_2$. **Figure S6a** demonstrates the $\kappa_{IP}$ of twisted bilayer MoS$_2$ with heat current flowing along the *x* and *y* directions for two typical sample sizes of 5 nm and 30 nm. In both *x* and *y* directions, a very weak dependence of $\kappa_{IP}$ on the twist angle is observed. In addition, the in-plane anisotropy ratio $\kappa_x/\kappa_y$ is close to ~1 and almost independent with the twist angle (**Figure S6b**), which indicates that the in-plane thermal transport properties of twisted bilayer MoS$_2$ is isotropic.

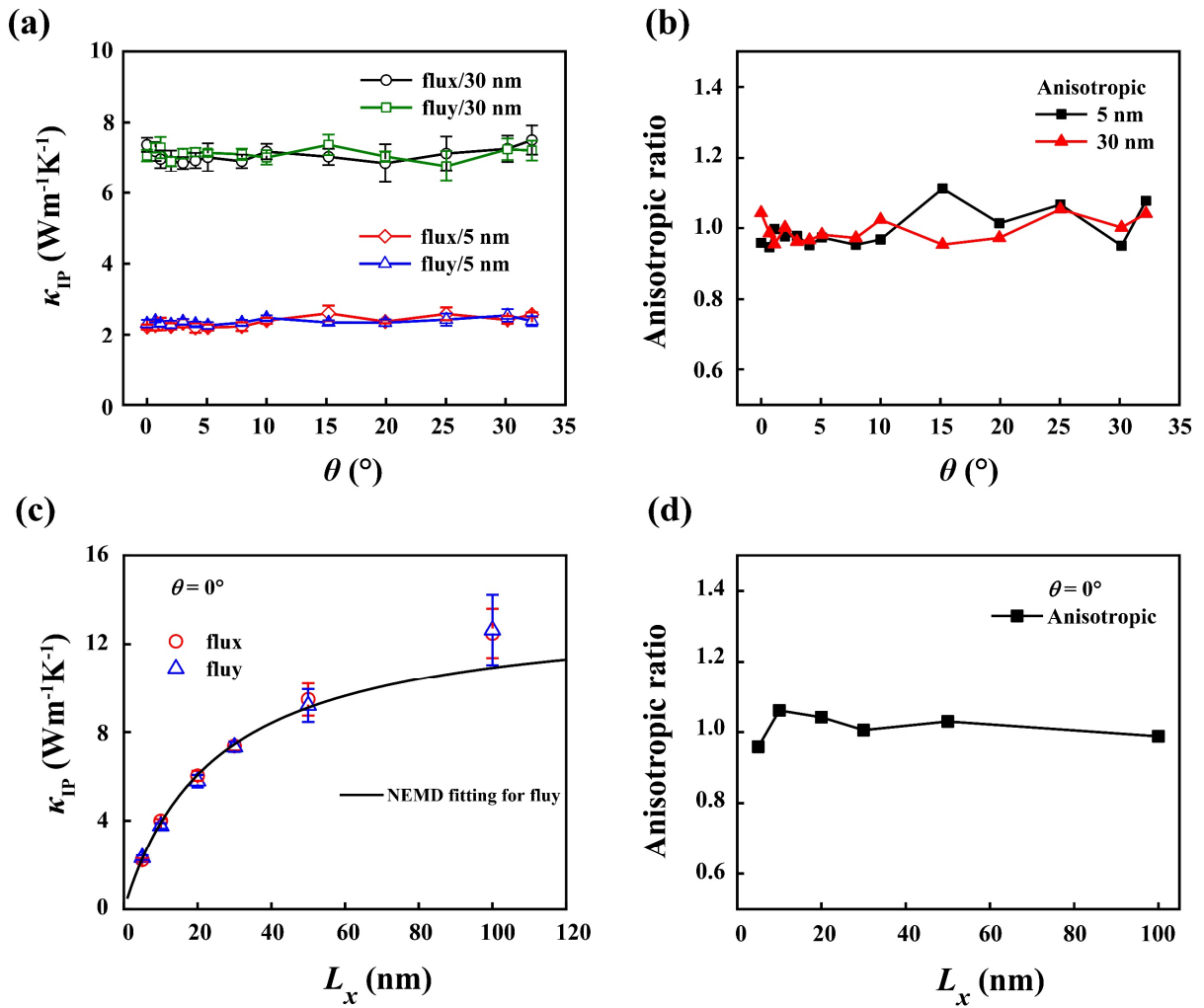

**Figure S6.** (a) The in-plane thermal conductivity ($\kappa_{IP}$) of bilayer MoS$_2$ as a function of the twist angle. Black circles and red diamonds represent the results of twisted MoS$_2$ with the heat current flowing along the x direction, and green squares and blue triangles represent the results of twisted MoS$_2$ with the heat current flowing along the y direction. (b) Twist-angle dependence of in-plane thermal conductivity anisotropic ratio. (c) The length dependence of in-plane thermal conductivity for MoS$_2$ stacks with zero twist angle between adjacent layers. Black circles and red triangles represent the results of untwisted MoS$_2$ stacks with the heat current flowing along the zigzag ($x$) and armchair ($y$) directions of the MoS$_2$ lattice, respectively. The solid blue line is the extrapolated fit of NEMD simulations (red triangle) using Eq. (1) in the main text. (d) Length-dependence of in-plane thermal conductivity anisotropic ratio.

To further explore the size effect of the anisotropy ratio, we calculated the $\kappa_{\mathrm{IP}}$ for the studied lengths along the $x$- (zigzag) and $y$- (armchair) directions ($L$ = 5 nm, 10 nm, 20 nm, 30 nm, 50 nm, and 100 nm) with untwisted bilayer MoS$_2$ ($\theta = 0°$). As is illustrated in **Figure S6c**, small difference in the $\kappa_{\mathrm{IP}}$ in the $x$- and $y$-directions for the same size, which leads to a weak anisotropy (see **Figure S6d**). Here, we also calculated the $\kappa_{\infty}$ with heat current flowing along the armchair direction using the most straightforward extrapolation equation (Eq. (1) in the main text),[7] and a linear fit to the NEMD data gives $\kappa_{\infty}$ = 13.6 ± 0.8 Wm$^{-1}$K$^{-1}$ and $\lambda$ = 24.9 ± 2.4 nm.

## 6. Temperature profiles

**Figure S7** demonstrates the temperature profiles extracted from the NEMD simulations for multilayer and bilayer MoS$_2$ systems to calculate the cross-plane ($\kappa_{CP}$) and in-plane ($\kappa_{IP}$) thermal conductivities. Note that only the linear regions of the temperature profiles in **Figure S7** were used to calculate $\kappa_{CP}$ and $\kappa_{IP}$.

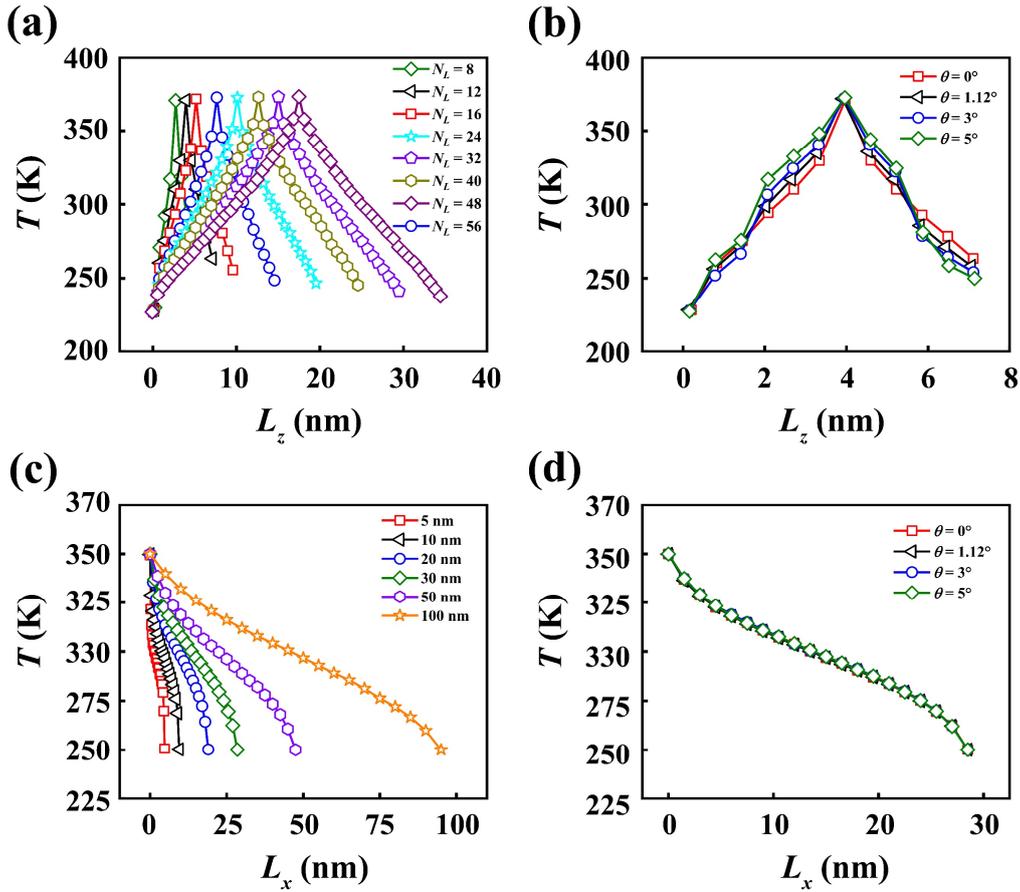

**Figure S7.** Temperature profiles for multilayer and bilayer MoS$_2$ systems. (a) Cross-plane temperature of the aligned MoS$_2$ system ($\theta = 0°$) with different stack thicknesses. $N_L$ represents the number of layers in the systems. (b) Cross-plane temperature of the twisted 12-layers MoS$_2$ system with several typical misfit angles ($\theta = 0°$, 1.12°, 3°, and 5°). (c) In-plane temperature of the aligned bilayer MoS$_2$ system ($\theta = 0°$) with different lengths along the heat flux direction ($L = 5$ nm, 10 nm, 20 nm, 30 nm, 50 nm, and 100 nm). (d) In-plane temperature of the twisted bilayer MoS$_2$ system with several typical misfit angles ($\theta = 0°$, 1.12°, 3°, and 5°).